**Exact solution of a two-dimensional Ising model with the next nearest interactions**

Zhidong Zhang

Shenyang National Laboratory for Materials Science, Institute of Metal Research, Chinese Academy of Sciences, 72 Wenhua Road, Shenyang, 110016, P.R. China

## Abstract

The exact solution of a two-dimensional (2D) Ising model with the next nearest interactions at zero magnetic field is derived. At first, the transfer matrices are analyzed in three representations, *i.e.,* Clifford algebraic representation, transfer tensor representation and schematic representation, to inspect nontrivial topological structures in this system. The system is equivalent to a triangular Ising model plus an interaction along the *z* axis, so that the approaches developed for the 3D Ising model are modified to be appropriable for solving the exact solution of the 2D Ising model with the next nearest interactions. The partition function and the spontaneous magnetization are obtained. The comparison with the exact solutions of other Ising lattices reveals that either the increase of the number of interactions in a unit cell or the presence/increase of topological contributions enhances the critical point of the Ising lattices. The results obtained in this work are helpful for understanding the physical properties of the 2D magnetic materials.

The corresponding author: Z.D. Zhang, e-mail address: zdzhang@imr.ac.cn

## 1.Introduction

The Ising model is one of the most fundamental models in physics, describing many-body spin (or particle) interactions [1-3]. Onsager [3] solved exactly the partition function, the free energy and the specific heat of the two-dimensional (2D) rectangular Ising model. Kaufman [4] developed a spinor representation to simplify the Onsager's procedure. Houtapple [5] and Newell [6] studied order-disorder phase transitions in hexagonal and triangular Ising lattices. Yang [7], Chang [8] and Potts [9] derived the exact solutions of the spontaneous magnetization of square, rectangular and triangular Ising lattices.

The exact solutions of the three-dimensional (3D) Ising model at zero magnetic field, the 2D Ising model at an external magnetic field and the 2D Ising model with the next nearest interactions at zero magnetic field are three well-known hard problems in the same difficulty level. Newell and Montroll [10] analyzed the theory of Ising models and pointed out that the difficulties for these three problems are topological. One meets serious hinders as one attempts to apply the algebraic method used for the 2D model at the absence of a magnetic field to these problems. The operators of interest generate a much large Lie algebra that it would be of little value, while nontrivial topological structures emerge. The procedures developed by Onsager [3], Kaufman [4], Kac and Ward [11] cannot be generated directly to be applicable for the three problems, which introduce some troubles in topology by counting closed graphs. However, the topological problems are different for the three problems. Any approaches based on only local environments cannot be exact for these problems.

In order to solve the ferromagnetic 3D Ising model at zero magnetic field, the present author proposed two conjectures in [12], investigated its mathematical structure in [13], then proved rigorously the two conjectures by a Clifford algebraic approach in collaboration with Suzuki and March [14], and further by a method of Riemann-Hilbert problem in collaboration with Suzuki [15,16]. Furthermore, the exact solutions of the 2D Ising model with a transverse field [17], the 3D $Z_2$ lattice gauge theory [18], the 3D spinless fermionic model [19] and the (3+1)-dimensional $\phi^4$ scalar field model with ultraviolet cutoff [20] were derived by the equivalence/duality between these models. Based on these results, topological quantum statistical mechanics and topological quantum field theories were investigated systematically [21]. The experimental data confirm the existence of the 3D Ising universality class [22,23], which affirm the validity of the exact solutions of the 3D Ising models [12]. The Monte Carlo simulations [24,25] on the critical exponents of the 3D Ising model, which were obtained by taking into account the nontrivial topological contributions of spin chains, agree well with our exact solutions. In addition, guided by a better understanding on the nontrivial topological structures in the 3D Ising models [12-16], the present author determined the lower bound of computational complexity of several NP-complete problems, such as spin-glass 3D Ising models [26], Boolean satisfiability problems [27], knapsack problems [28] and traveling salesman problems [29]. In a recent work [30], the exact solution of the 2D rectangular Ising model at an external magnetic field is derived by a modified Clifford algebraic approach.

The motivations of the present work are as follows: The exact solution of the 2D Ising model with the next nearest interactions is a long-standing unsolved problem in physics. The approaches developed for solving exactly the 3D Ising model at the absence of a magnetic field may guide the route to derive the exact solution of the 2D Ising model with the next nearest interactions. The exact solution is extremely important for inspecting physical properties of the 2D Ising models as well as the 2D magnetic materials.

This paper is arranged along the following line of presentation: In Section 2, the model is set up and the transfer matrices are described respectively by a Clifford algebraic representation, a transfer tensor representation and a schematic representation, to inspect nontrivial topological structures. In Section 3, the partition function and the magnetization of the 2D rectangular Ising model with the next nearest interactions are derived explicitly by a Clifford algebraic approach. Section 4 is for conclusions.

**2. Model and transfer matrices**

The Hamiltonian of the 2D Ising model with the next nearest interactions at zero magnetic field is written as:

$$\hat{\mathrm{H}} = -\sum_{<i,j>}^{n,m} \left[ J_1 s_{i,j} s_{i+1,j} + J_2 s_{i,j} s_{i,j+1} + J_3 s_{i+1,j} s_{i,j+1} + J_4 s_{i,j} s_{i+1,j+1} \right] \tag{1}$$

Here every Ising spin $s_{i,j} = \pm 1$ for spin up and spin down is located on a rectangular lattice with the lattice size $N = mn$. The numbers $(i, j)$ run from (1, 1) to $(m, n)$, denoting lattice points along two crystallographic directions. The nearest neighboring interactions $J_1$ and $J_2$, and the next nearest interactions $J_3$ and $J_4$ are considered, which are all ferromagnetic, and thus no frustration presents. One may set $J_1 \geq J_2$ and $J_3 \geq J_4$ without loss of generality. If either $J_3 = 0$ or $J_4 = 0$, the system will be a triangular Ising model.

The spins at $(i, j)$, $(i+1, j)$ and $(i, j+1)$ can be represented as $(i, j, 1)$, $(i+1, j, 1)$ and $(i, j+1, 1)$ for the first layer. The spin at $(i+1, j+1)$ can be denoted as $(i, j, 2)$, as viewed to be at the second layer. The Hamiltonian can be rewritten as:

$$\widehat{\mathrm{H}} = -\sum_{<i,j>}^{n,m} \left[ J_1 s_{i,j,1} s_{i+1,j,1} + J_2 s_{i,j,1} s_{i,j+1,1} + J_3 s_{i+1,j,1} s_{i,j+1,1} + J_4 s_{i,j,1} s_{i,j,2} \right] \quad (2)$$

The Hamiltonian (2) represents the 2D Ising model on a triangular lattice (the first layer) with an interaction $J_4$ along the $z$ direction. The system becomes the triangular Ising model with spins interacting with the spins in the neighboring layer (the second layer). The interaction $J_4$ is acting as the interaction along the third crystallographic direction in the 3D Ising model, which results in long-range spin entanglements and nontrivial topological structures. The Hamiltonians (1) and (2) are equivalent.

2.1. Clifford algebraic representation

The partition function $Z$ of the 2D Ising model with the next nearest interactions is expressed as follows:

$$Z = (2sinh2K_1)^{\frac{n}{2}} \cdot \mathrm{trace}(\boldsymbol{V})^m \equiv (2sinh2K_1)^{\frac{n}{2}} \cdot \sum_{i=1}^{2^n} \lambda_i^m \quad (3)$$

with the transfer matrix $\boldsymbol{V} = \boldsymbol{V_4}\boldsymbol{V_3}\boldsymbol{V_2}\boldsymbol{V_1}$ as:

$$\boldsymbol{V_4} = \prod_{j=1}^{n} exp\left[ -iK_4 \Gamma_{2j} \left( \prod_{k=j+1}^{j+n-1} i\,\Gamma_{2k-1}\Gamma_{2k} \right) \Gamma_{2j+2n-1} \right] \quad (4)$$

$$\boldsymbol{V_3} = \prod_{j=1}^{n} exp\left[-iK_3\Gamma_{2j}\Gamma_{2j+1}\right] \tag{5}$$

$$\boldsymbol{V_2} = \prod_{j=1}^{n} exp\left[-iK_2\Gamma_{2j}\Gamma_{2j+1}\right] \tag{6}$$

$$\boldsymbol{V_1} = \prod_{j=1}^{n} \exp\left[iK_1^*\Gamma_{2j-1}\Gamma_{2j}\right] \tag{7}$$

It is convenient to introduce variables $K_l \equiv J_l/(k_B T)$ instead of interactions $J_l$, ($l$ = 1, 2, 3, 4). The Kramers-Wannier relation defines the dual interaction $K_1^* = \frac{1}{2} ln(cothK_1)$ [2]. The generators of Clifford algebra are written as:

$$\Gamma_{2j-1} \equiv P_j = \mathrm{C} \otimes \mathrm{C}\otimes \ldots \otimes\mathrm{C}\otimes\mathrm{s}'\otimes\mathrm{I}\otimes\ldots\otimes\mathrm{I} \quad (j-1 \;\; times \;\; C) \tag{8}$$

$$\Gamma_{2j} \equiv Q_j = \mathrm{C} \otimes \mathrm{C}\otimes \ldots \otimes\mathrm{C}\otimes(-is'')\otimes\mathrm{I}\otimes\ldots\otimes\mathrm{I} \quad (j-1 \;\; times \;\; C) \tag{9}$$

The $\Gamma_{2j-1}$ and $\Gamma_{2j}$ (and also $P_j$ and $Q_j$) matrices are referred to the Onsager-Kaufman-Zhang notations [3,4,12-16]. $\mathrm{s}'' = \begin{bmatrix} 0 & -1 \\ 1 & 0 \end{bmatrix}$ $(= -i\sigma_2)$, $\mathrm{s}' = \begin{bmatrix} 1 & 0 \\ 0 & -1 \end{bmatrix}$ $(= \sigma_3)$, $C = \begin{bmatrix} 0 & 1 \\ 1 & 0 \end{bmatrix}$ $(= \sigma_1)$, where $\sigma_j$ ($j$ = 1,2,3) are Pauli matrices, while $\boldsymbol{I}$ is the unit matrix. In the transfer matrices, the boundary factor $\boldsymbol{U}$ in Kaufman's paper [4] is neglected, since it splits the space into two subspaces, and in the thermodynamic limit the surface to volume ration vanishes for an infinite system according to the Bogoliubov inequality [14].

The linear terms in the transfer matrices $\boldsymbol{V_1}$, $\boldsymbol{V_2}$ and $\boldsymbol{V_3}$ are the same as those for the triangular Ising lattice [5,6,9]. The nonlinear terms in the transfer matrix $\boldsymbol{V_4}$ are induced by the interaction $J_4$, which cause the topological problems including nonlocality, nonlinearity, non-commutative and non-Gaussian. Similar to the 3D case [14], these problems are roots of difficulties hindering exactly solving the problems. On one hand, the 2D Ising model with the next nearest interactions exhibits a topological problem, being the same as that in the 3D Ising model at zero magnetic field. We can employ the Clifford algebraic approach developed in [14] with some modifications. An additional rotation is added to trivialize the nontrivial topological structures to be diagonalizable, while topological phases are generalized on eigenvectors and eigenvalues. The only difference is that the dimensionalities in the two models are different.

2.2. Transfer tensor representation

The partition function $Z$ of the 2D Ising model with the next nearest interactions can be represented also in the transfer tensor forms:

$$Z = (2sinh2K_1)^{\frac{n}{2}} \cdot \mathrm{trace}(\boldsymbol{T})^m \tag{10}$$

with the transfer tensor $\boldsymbol{T}$ as:

$$\boldsymbol{T} = \prod_{j=1}^{n} exp\left[T_{uvwt}^{(j)}\right] \tag{11}$$

The basic elements in the transfer tensor representation can be expressed by:

$$T_{uvwt} = \langle s_{i,j}, s_{i+1,j}, s_{i,j+1}, s_{i+1,j+1} \left| e^{-\hat{H}/k_B T} \right| s_{i,j}, s_{i+1,j}, s_{i,j+1}, s_{i+1,j+1} \rangle \tag{12}$$

All possible combinations of four Ising spins with values ± 1 give sixteen configurations, which can be described by a four-order tensor $T_{uvwt}$ with $u$, $v$, $w$, $t$ = 1, 2. as follows:

$$T_{1111} = \langle + + + + \left| e^{-\hat{H}/k_B T} \right| + + + + \rangle = e^{K_1^* + K_2 + K_3 + K_4}, \quad (13)$$

$$T_{2111} = \langle - + + + \left| e^{-\hat{H}/k_B T} \right| - + + + \rangle = e^{-K_1^* - K_2 + K_3 - K_4}, \quad (14)$$

$$T_{1211} = \langle + - + + \left| e^{-\hat{H}/k_B T} \right| + - + + \rangle = e^{-K_1^* + K_2 - K_3 + K_4}, \quad (15)$$

$$T_{2211} = \langle - - + + \left| e^{-\hat{H}/k_B T} \right| - - + + \rangle = e^{K_1^* - K_2 - K_3 - K_4}, \quad (16)$$

$$T_{1121} = \langle + + - + \left| e^{-\hat{H}/k_B T} \right| + + - + \rangle = e^{K_1^* - K_2 - K_3 + K_4}, \quad (17)$$

$$T_{2121} = \langle - + - + \left| e^{-\hat{H}/k_B T} \right| - + - + \rangle = e^{-K_1^* + K_2 - K_3 - K_4}, \quad (18)$$

$$T_{1221} = \langle + - - + \left| e^{-\hat{H}/k_B T} \right| + - - + \rangle = e^{-K_1^* - K_2 + K_3 + K_4}, \quad (19)$$

$$T_{2221} = \langle - - - + \left| e^{-\hat{H}/k_B T} \right| - - - + \rangle = e^{K_1^* + K_2 + K_3 - K_4}, \quad (20)$$

$$T_{1112} = \langle + + + - \left| e^{-\hat{H}/k_B T} \right| + + + - \rangle = e^{K_1^* + K_2 + K_3 - K_4}, \quad (21)$$

$$T_{2112} = \langle - + + - \left| e^{-\hat{H}/k_B T} \right| - + + - \rangle = e^{-K_1^* - K_2 + K_3 + K_4}, \quad (22)$$

$$T_{1212} = \langle + - + - \left| e^{-\hat{H}/k_B T} \right| + - + - \rangle = e^{-K_1^* + K_2 - K_3 - K_4}, \quad (23)$$

$$T_{2212} = \langle - - + - \left| e^{-\hat{H}/k_B T} \right| - - + - \rangle = e^{K_1^* - K_2 - K_3 + K_4}, \quad (24)$$

$$T_{1122} = \langle + + - - \left| e^{-\hat{H}/k_B T} \right| + + - - \rangle = e^{K_1^* - K_2 - K_3 - K_4}, \quad (25)$$

$$T_{2122} = \langle - + - - \left| e^{-\hat{H}/k_B T} \right| - + - - \rangle = e^{-K_1^* + K_2 - K_3 + K_4}, \quad (26)$$

$$T_{1222} = \langle + - - - \left| e^{-\hat{H}/k_B T} \right| + - - - \rangle = e^{-K_1^* - K_2 + K_3 - K_4}, \quad (27)$$

$$T_{2222} = \langle - - - - \left| e^{-\hat{H}/k_B T} \right| - - - - \rangle = e^{K_1^* + K_2 + K_3 + K_4}, \quad (28)$$

The transfer tensors of the whole system can be represented by the direct products of the transfer tensors for all the spins in the 2D Ising model. The transfer tensor representation illustrates concisely all the terms of the Hamiltonian in a stereoscopic form. The nontrivial topological structures are hidden in the hypercubic formulates, since the hypercubic matrix has the 3D character that fits well with the 3D character of the basic elements for the 2D Ising model with the next nearest interactions. It is clear that the states of spins in the present model are equivalent to the states of a triangular Ising lattice with an additional interaction along the third direction. However, the diagonalization in the transfer tensor representation cannot derive directly the desired solution, because the Lie algebra is so small that the number of the body-diagonalization elements are not enough for eigenvalues of the system. Meanwhile, the corresponding eigenvectors are not large enough for representing the Hilbert space of the system. For exactly solving the problem, the transfer tensors have to be mapped into the transfer matrices for a diagonalization. Nevertheless, the transfer tensor representation confirms the equivalence between the Hamiltonians (1) and (2), and gives some implications on the Clifford algebraic approach for solving the problem.

2.3. Schematic representation

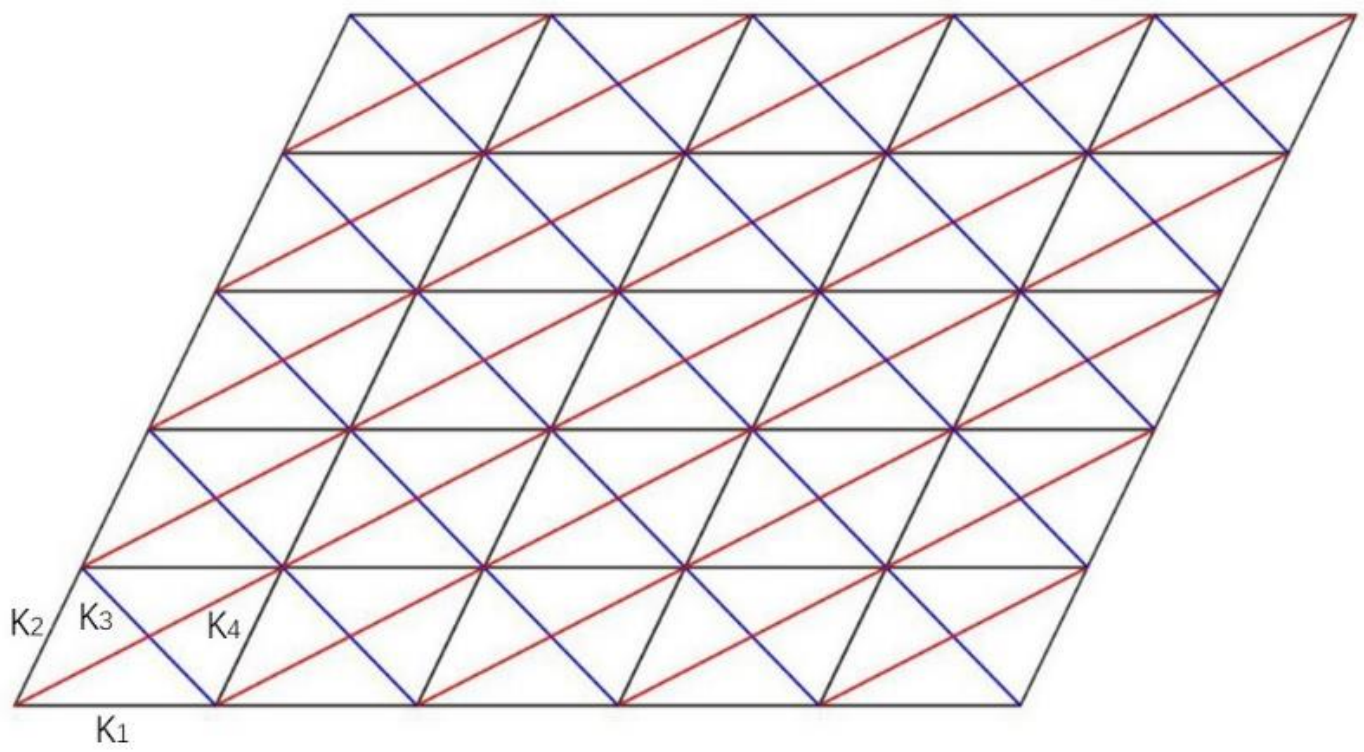

Figure 1. Illustration of a rectangular Ising model with the next nearest interactions on a 6×6 lattice. The nearest neighboring interactions $K_1$ and $K_2$ are represented by black lines, while the next nearest interactions $K_3$ and $K_4$ are represented by blue and red lines respectively.

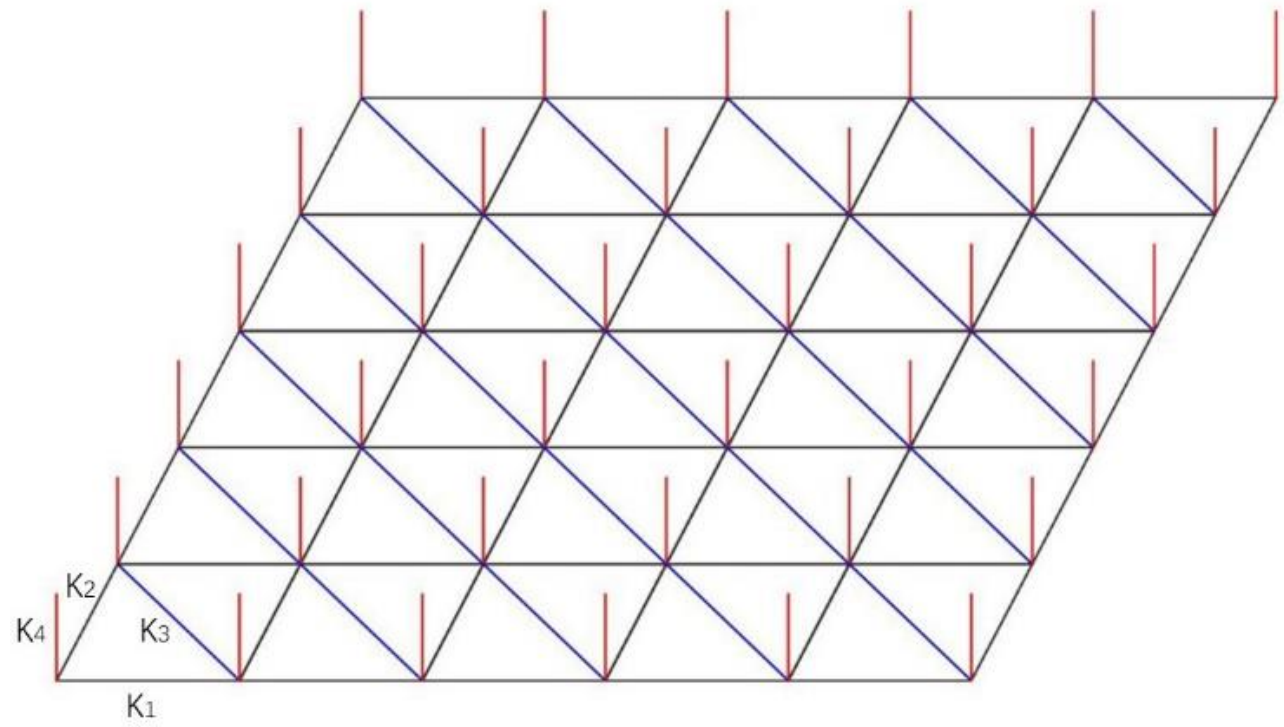

Figure 2. Illustration of a triangular Ising model with interactions $K_1$, $K_2$ and $K_3$ on a 6×6 lattice plus an interaction $K_4$ along the $z$ direction at each lattice point. The interactions $K_1$ and $K_2$ are represented by black lines, the interaction $K_3$ is represented by blue lines, and the interaction $K_4$ is represented by red lines.

We utilize the schematic representation to illustrate the topological structures of the 2D Ising model with the next nearest interactions, demonstrating the transformation from one topological state to another, which keeps the equivalence of the free energy of the system. Figure 1 shows schematically a rectangular Ising model with the nearest neighboring interactions $K_1$ and $K_2$, and the next nearest interactions $K_3$ and $K_4$ on a 6×6 lattice (representing the Hamiltonian (1)). Figure 2 illustrate a topological structure of a triangular Ising model with interactions $K_1$, $K_2$ and $K_3$ on a 6×6 lattice plus an interaction $K_4$ along the $z$ direction at each lattice point (representing the Hamiltonian (2)). Topologically, the structures in Figures 1 and 2 are equivalent. In previous work [20-23] for determining the lower bound of the computational complexity of NP-complete problems, an absolute minimum core (AMC) model in the spin-glass 3D Ising model was defined as a spin-glass 2D Ising model interacting with its nearest neighboring plane. The 2D Ising model with the next nearest interactions has the same structure as the AMC model, but now the interactions within the 2D lattice are all ferromagnetic and the next nearest interaction $K_4$ is acting along the third crystallographic direction. Notice that in Figures 1 and 2, as an example, only are the 2D lattice with finite lattice points illustrated. In the thermodynamic limit, the number $N = nm$ of the lattice points approaches infinite (*i.e.*, $n \rightarrow \infty$, $m \rightarrow \infty$, $N \rightarrow \infty$). The nontrivial topological structures illustrated in Figure 2 are consistent with those hidden in the Clifford algebraic formulas (Eqs. (4)-(7)) for the transfer matrices. This fact strongly suggests that the approaches developed for the 3D Ising model can be employed for solving the 2D Ising model with the next nearest interactions.

## 3. Partition function and spontaneous magnetization

### 3.1. Clifford algebraic approach

The Clifford algebraic approach developed for solving exactly the ferromagnetic 3D Ising model in zero magnetic field is described briefly as follows [14]: According to the topology theory [31], the crosses of knots/links in the topological structures contribute also to the partition function of a physical system. There is a mapping between a cross and a spin, and in a 3D Ising model two contributions consist of local spin alignments and nonlocal spin entanglements [16]. By performing a time average in the Jordan-von Neumann-Wigner framework of the quantum mechanics [32], one can deal with the non-commutation of operators. By using some basic facts of the direct product and the trace, the 3D Ising model is extended to be (3+1)-dimensional, and then divided to many sub-models with sub-transfer matrices in the quasi-2D limit. This process overcomes difficulties (such as nonlocality, nonlinearity, non-commutative and non-Gaussian) for solving the problem. By employing the Kaufman's procedure for the 2D Ising model [4], respecting with the same character of the internal factor $W_j$ and the boundary factor $U$, nonlinear terms in the transfer matrices are linearized while the Hilbert spaces are splitting. By introducing a topological Lorentz transformation, which is seen as a local gauge transformation, the 3D Ising model can be transferred from a nontrivial topological basis to a trivial topological basis, while generalizing the topological phases and taking into account the contribution of the nontrivial topological structures to the partition function and the thermodynamic properties. Finally, the desired solution is realized for the 3D Ising model by fixing the rotation angle for the local gauge transformation and the phase factors [12-16]. The Clifford algebraic approach can be modified to be appropriate for deriving its exact solution of the present system. An additional rotation for the topological Lorentz transformation is added so that the dimension of the model is extended to be (2+1)-dimensional with two topological phases. The rotation angle is determined by the star-triangle relation (*i.e,* Yang-Baxter equation).

3.2. Partition function

According to Onsager [3] and Kaufman [4], the planar rotations in the spinor representation can be transformed into the rotation representation. the eigenvalues and the partition function can be calculated by the planar rotations in the rotation representation. Following the procedure in our previous work [12-16], the procedure for solving exactly the 2D Ising model with the next nearest interactions is straightforward. The partition function of the triangular Ising lattice obtained by Houtapple [5] can be generalized by employing the Clifford algebraic approach.

The partition function of the 2D rectangular Ising model with the next nearest interactions is represented as:

$$N^{-1} lnZ = ln2$$
$$+ \frac{1}{2(2\pi)^3} \int_0^{2\pi} \int_0^{2\pi} \int_0^{2\pi} ln\{cosh2K_1 cosh2(K_2 + K_4 + K_5) cosh2K_3$$
$$+ sinh2K_1 sinh2(K_2 + K_4 + K_5) sinh2K_3 - sinh2K_1 cos\omega_1$$
$$- sinh2(K_2 + K_4 + K_5)[cos\omega_2 cos\phi + cos\omega'_2 cos\phi']$$
$$- sinh2K_3[cos(\omega_1 + \omega_2) cos\phi$$
$$+ cos(\omega_1 + \omega'_2) cos\phi']\} \, d\omega_1 d\omega_2 d\omega'_2$$

(29)

Here the interaction $K_5 = \frac{K_2 K_4}{K_1}$ corresponds to the rotation angle for the Lorentz transformation (*i.e.* the local gauge transformation). The rotation angle $K_5$ is determined by the star-triangle relation $K_1 K_1^* = K_1 K_2 + K_1 K_4 + K_2 K_4$ . The topological phases $\phi$ and $\phi'$ at finite temperatures equal to 2π and π/2, respectively, similar to the 3D case [12-16]. The topological phases are connected with the Gauss-Bonnet-Chern formula [33], the Röhrl theorem for a monodromy representation [34], the Aharonov-Bohm effect [35], the Berry phase [36], *etc.* (see [13-16,18,30] for detail). In Eq. (29), when $K_3 \neq 0$, $K_4 = 0$, one has $K_5 = 0$, the solutions return to the triangular Ising model [5]. Note that the critical point of the triangular Ising model with the same interactions $K_1 = K_2 = K_3$ is located at $x_c = e^{-2K_c} = \frac{1}{\sqrt{3}}$ , $\frac{1}{K_c} = 3.6409569 \ldots$ From the partition function in Eq. (29), we can calculate the free energy and the specific heat, and obtain the critical exponent α = 0 for the specific heat.

At the beginning of the procedure, we have pre-set the conditions $J_1 \geq J_2$ and $J_3 \geq J_4$. The partition function (Eq. (10)) is held only for such conditions. For $J_1 \leq J_2$, the parameters $K_1$ and $K_2$ should be permutated. Since the present problem is similar to the 3D simple orthorhombic Ising lattices, the reason is analogous to the 3D case. This is because the star-triangle relations would lead to three possible solutions for the critical point, but only one of them is the real solution of the system. One has to choose the one for the lowest critical point, in consideration of the symmetry of the system. At the beginning of the diagonalization procedure, we have to set up only the largest (say $K_1$) among the interactions as the standard axis for the definition of the interaction $K_1^*$ in the dual lattice, because the solution obtained in this way is the lowest. The simple cubic lattice with the highest symmetry has the highest critical temperature, at the golden ratio. Any solution higher than the golden ratio is forbidden for the 3D simple orthorhombic Ising lattices due to lack of physical significance (for detail discussion, readers refer to pages 5395-5396 of [12]). For $J_3 \leq J_4$, the parameters $K_3$ and $K_4$ should be permutated, in order to minimize the topological contributions to the critical point.

3.3. Spontaneous magnetization

The perturbation procedure with a weak virtual magnetic field $\aleph$ developed by Yang [7] can be generated to derive the spontaneous magnetization of the 2D Ising model with the next nearest interactions. The procedure is straightforward with replacements of parameters in the formula for the triangular Ising lattice obtained by Potts [9], in consideration of the interaction $K_4$ and the additional rotation $K_5$ for trivializing the nontrivial topological structures.

The spontaneous magnetization of the 2D rectangular Ising model with the next nearest interactions is represented as:

$$M = \left[1 - \frac{16x_1^2x_2^2x_3^2x_4^2x_5^2}{(1 + x_1x_2x_4x_5 + x_2x_3x_4x_5 + x_1x_3)(1 + x_1x_2x_4x_5 - x_2x_3x_4x_5 - x_1x_3)} \times \frac{1}{(1 - x_1x_2x_4x_5 + x_2x_3x_4x_5 - x_1x_3)(1 - x_1x_2x_4x_5 - x_2x_3x_4x_5 + x_1x_3)}\right]^{1/8} \tag{30}$$

with $x_l = e^{-2K_l}$ ($l$ =1, 2, 3, 4, 5). Note the critical exponent β equals to 1/8, since the system keeps the 2D nature, while the AMC model has a quai-2D character. Again, the spontaneous magnetization (Eq.(11)) is validated only for the conditions $J_1 \geq J_2$ and $J_3 \geq J_4$. For $J_1 \leq J_2$, the parameters $K_1$ and $K_2$ should be permutated. For $J_3 \leq J_4$, the parameters $K_3$ and $K_4$ should be permutated

For a particular case, when $K_1 = K_2$ for a square lattice and $K_3 = K_4$, namely $x_1 = x_2$ and $x_3 = x_4 = x_5$, one has:.

$$M = \left[1 - \frac{16x_1^4x_3^6}{(1 + x_1^2x_3^2 + x_1x_3^3 + x_1x_3)(1 + x_1^2x_3^2 - x_1x_3^3 - x_1x_3)} \times \frac{1}{(1 - x_1^2x_3^2 + x_1x_3^3 - x_1x_3)(1 - x_1^2x_3^2 - x_1x_3^3 + x_1x_3)}\right]^{1/8} \tag{31}$$

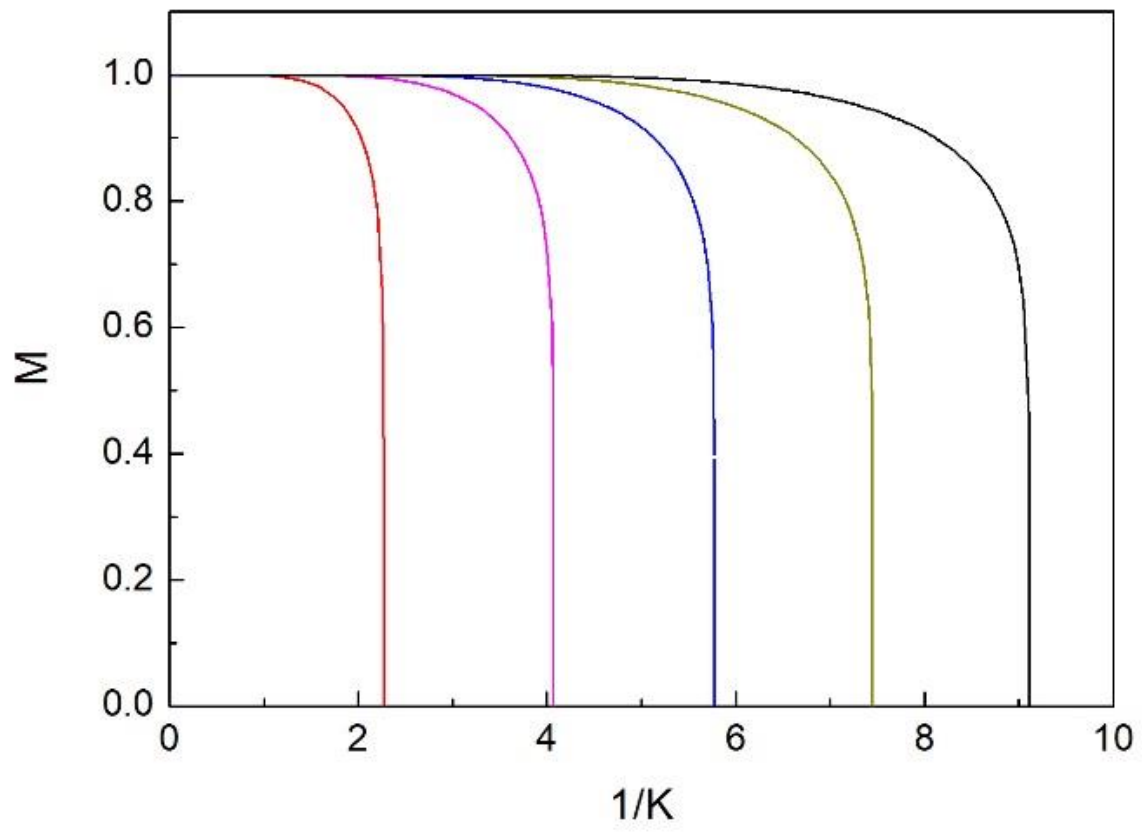

Figure 3. Temperature dependence of the spontaneous magnetization of the 2D Ising

model with the next nearest interactions when $K_1 = K_2$ and $K_3 = K_4$. The curves from left to right correspond to the next nearest interaction $K_3 = 0$, $0.5K_1$, $K_1$, $1.5K_1$ and $2K_1$, respectively.

Figure 3 represents the temperature dependence of the spontaneous magnetization of the 2D Ising model with the next nearest interactions when $K_1 = K_2$ and $K_3 = K_4$. At a fixed temperature, the spontaneous magnetization is enhanced upon the increase of the next nearest interactions. The critical point is located at the silver solution $x_c = e^{-2K_c} = \sqrt{2} - 1$, $\frac{1}{K_c} = 2.26918531 \ldots$ for the square Ising lattice without the next nearest interactions ($K_3 = K_4 = 0$). The increase of the next nearest interactions $K_3$ and $K_4$ increases the critical point for the order-disorder phase transition. For $K_3 = K_4 = 0.5K_1$, the critical point is located at about $\frac{1}{K_c} = 4.069257$, which is already higher than the critical point $\frac{1}{K_c} = 3.6409569 \ldots$ for the triangular Ising model. Such an increase is ascribed to the contribution of the nontrivial topological structures. For the simplest case, when all the interactions equal $K_1 = K_2 = K_3 = K_4 = K$, namely $x_1 = x_2 = x_3 = x_4 = x_5 = x = e^{-2K}$, one has:.

$$M = \left[1 - \frac{16x^{10}}{(1-x^2)^2(1+x^2+2x^4)(1+x^2-2x^4)}\right]^{1/8} \tag{32}$$

The critical point is located at about $\frac{1}{K_c} = 5.77078017 \ldots$. Then, the next nearest interactions increase the critical point to $\frac{1}{K_c} \approx 7.4441673 \ldots$ for $K_3 = K_4 = 1.5K_1$ and to $\frac{1}{K_c} \approx 9.10541124 \ldots$ for $K_3 = K_4 = 2K_1$.

Table 1. Comparisons of the critical point, the number of interactions in a unit cell, topological contributions in a unit cell for several Ising lattices.

| lattice | Critical point $\frac{1}{K_c}$ | Number of interactions in a unit cell | Topological contributions in a unit cell |
|---|---|---|---|
| Square lattice | 2.26918531 ... | 2 | 0 |
| Triangular lattice | 3.6409569 ... | 3 | 0 |
| Square lattice with $K_3 = K_4 = 0.5K$ | 4.069257 ... | 3 | $0.5K$ |
| Cubic lattice | 4.15617384 .. | 3 | $K$ |
| Square lattice with $K_3 = K_4 = K$ | 5.77078017 ... | 4 | $K$ |
| Hypercube lattice | 8 | 4 | $\infty$ |

It is interesting to compare the critical points of different Ising lattices in order to reveal physical significance underlying phenomena. Table 1 represents the comparisons of the critical point, the number of interactions in a unit cell, topological contributions for several Ising lattices. The critical point increases in the following sequence: the square lattice, the triangular lattice, the square lattice ($K_1 = K_2$) with the next nearest interactions $K_3 = K_4 = 0.5K_1$, the 3D cubic lattice, the square lattice ($K_1 = K_2$) with the next nearest interactions $K_3 = K_4 = K_1$ and the hypercubic lattice in 4D (see Table 1). The critical point for the ferromagnetic - paramagnetic phase transition is located at the golden ratio $x_c = e^{-2K_c} = \frac{\sqrt{5}-1}{2}$, $\frac{1}{K_c} = 4.15617384$ ... for the cubic Ising lattice [12]. The interactions and the thermal activation energy balance at the golden ratio for the most symmetrical lattice in three dimensions. Note that the hypercubic lattice can be constructed by infinite number of cubic lattices so that there are infinite multiples of topological contributions in a cubic lattice. The critical point of the hypercubic lattice is determined by the mean-field theory to be 8 (the coordination number). It is clear from Table 1 that either the increase of the number of interactions in a unit cell or the presence/increase of topological contributions enhances the critical point of the Ising lattices.

## 4. Conclusions

In conclusion, the exact solution of the 2D Ising model with the next nearest interactions at zero magnetic field is derived by a Clifford algebraic approach. Inspecting on the transfer matrices by the Clifford algebraic representation, the transfer tensor representation and the schematic representation confirms the existence of the nontrivial topological structures in this system. A topological Lorentz transformation is applied for dealing with the topological problem in the present system. The rotation angle for the transformation is determined by the star-triangular relation. The partition function and the spontaneous magnetization are obtained analytically. It is found that either the increase of the number of interactions in a unit cell or the presence/increase of topological contributions enhances the critical point of the Ising lattices. Solving exactly the Ising models not only understands in-depth these models themselves [12-21], but also benefit to solving the hard problems in mathematics and computer sciences [26-29,37,38].

**Acknowledgements**

This work has been supported by the National Natural Science Foundation of China under grant number 52031014.

**Data availability** Data available upon request from the author.

**Conflict of interest** The author declares that this contribution is no conflict of interest.

**Author contribution statements** Z.D. Zhang is the only author, who contribute to conception, method, investigation, validation, visibility and writing the manuscript.